\documentclass[12pt]{article}
\usepackage{epsfig}
\usepackage{amssymb}
\usepackage{pslatex}

\makeatletter

\@addtoreset{equation}{section}
\makeatother
\newcommand{\bea}{\begin{eqnarray}}
\newcommand{\eea}{\end{eqnarray}}
\newcommand{\be}{\begin{eqnarray}}
\newcommand{\ee}{\end{eqnarray}}

\def\fR{{\mathfrak R}}
\def\fL{{\mathfrak L}}
\def\fQ{{\mathfrak Q}}
\def\fS{{\mathfrak S}}

\begin{document}

\begin{titlepage}
\vskip1cm
\begin{flushright}
$\mathbb{UOSTP}$ {\tt 08121}
\end{flushright}
\vskip1.25cm
\centerline{\Large \bf  
Zero Modes for the Boundary Giant Magnons}
\vskip1.25cm
\centerline{\large  Dongsu Bak
}
\vspace{1.25cm}
\centerline{\sl  Physics Department, University of Seoul, Seoul 
130-743 {\rm KOREA}}
\vskip0.25cm
\centerline{\tt dsbak@uos.ac.kr }
\vspace{1.5cm}
\centerline{ABSTRACT}
\vspace{0.75cm}
\noindent
We study the fermion zero-mode dynamics for open strings ending 
on the giant graviton branes.
For the open string ending on the $Z=0$ brane, the quantization of the
 fermion zero-modes of boundary giant magnons 
reproduces the 256 states of the boundary degrees with the 
precise realization 
of the $SU(2|2)\times SU(2|2)$ symmetry algebra.
Also for the open string ending on the $Y=0$ brane, we reproduce
the unique vacuum state from  the fermion zero-modes.

\end{titlepage}

\section{Introduction}

There have been great advances in our understanding of the 
correspondence between 
the planar ${\cal N}=4$ 
super-Yang-Mills (SYM) theory and the type IIB strings 
on the AdS$_5\times S^5$ background\cite{Maldacena:1997re}. 
The integrability plays a crucial role
for the check of the correspondence of the string sigma model and
the SYM spin chain dynamics\cite{Minahan:2002ve, Beisert:2003tq,
Beisert:2003yb, Beisert:2005fw, Bena:2003wd}. 

The elementary excitation of the SYM 
spin chains,  which is called a magnon, is composed of
16 states that are organized by 
$SU(2|2)\times SU(2|2)$ symmetry.  The S matrix 
describing the scattering of  two magnons 
is fully determined  and becomes  the basis of solving the 
asymptotic 
spectrum of the SYM spin chain operators\cite{
Staudacher:2004tk, Beisert:2005tm, Beisert:2006qh}.
In the string sigma model side, the giant magnon solution
 of the spinning
string  describes the corresponding 
elementary excitation\cite{Hofman:2006xt}.
In Ref.~\cite{Minahan:2007gf}, the dynamics of
the fermion zero modes around the giant magnon solution was studied. 
It was shown that
the quantization of these zero modes is precisely reproducing the
16 states of the magnon of the SYM spin chain side.      

In Ref.~\cite{Hofman:2007xp}, the correspondence between open 
spin chain in the ${\cal N}=4$ SYM 
theory and the open strings ending on the giant gravitons  was proposed.
As we shall explain details later on, there are two classes
 of open SYM
spin chain operators: One is the so called open spin chain 
of the  $Z=0$ brane
and the other is  open spin chain of 
the $Y=0$ brane.
The $Z=0$ brane vacuum involves two boundary degrees, which are 
localized at the left and the right boundaries respectively. It was shown 
that each  boundary magnon is again organized by the
$SU(2|2)\times SU(2|2)$ symmetry carrying 16 states. Hence there are 256 states
in total. On the other hand,  the ground state for the
 open spin chain of the $Y=0$ brane does not involve any boundary 
states and is characterized by a unique vacuum state. 

The reflection 
amplitudes of bulk magnon on the boundary were constructed
up to  overall phases in Ref.~\cite{Hofman:2007xp}. 
The overall dressing phases for the $Y=0$ brane and the 
$Z=0$ brane were later 
determined respectively in Ref.~\cite{Chen} and Ref.~\cite{AhnB}.
For the related aspects of the open spin chain correspondence,
see Refs.~\cite{Beren}.

In this note, we shall first construct the finite-size boundary giant 
graviton solution of the strings ending on the  $Z=0$ giant graviton brane. 
We study  the fermion zero mode dynamics of the string sigma model 
and identify the 256 states that are 
organized by the $SU(2|2)\times SU(2|2)$ symmetry.
For the strings ending on the $Y=0$ brane, we also construct the finite-size
vacuum solution and show that the quantization of the fermion zero mode
leads to the unique vacuum state of no boundary degrees.

\section{Boundary  states }

In Ref.~\cite{Hofman:2007xp},
 it was proposed 
that
 a magnon in a class of open spin chain in ${\cal N}=4$
super Yang-Mills theory is corresponding to 
a configuration of open string
ending on a giant graviton.  We shall briefly review
the relevant part of this proposal here.

The proposal is an open spin chain and open string version of the
the closed string giant magnon dynamics~\cite{Hofman:2006xt}.
There are two types of giant gravitons that allow BPS ground-state
configuration. If the
giant graviton is located at $Y=0$ or $Z=0$ hyper surfaces inside
${S}^5$, they are called
$Y=0$ and $Z=0$ branes, respectively. We choose the open string vacuum oriented
along $Z$-direction.
We see that the open string can end on the $Y=0$ brane with Neumann boundary
condition.
The open string can also end
on the $Z=0$ brane with Dirichlet boundary condition;
An additional localized boundary degree is  necessary
 at each boundary of $Z=0$.

In the ${\cal N}=4$ SYM theory 
side, 
the $Y=0$ brane open spin chain is
represented by composite operators 
containing a determinant factor ${\rm det}(Y)$:
\be
{\cal O}_Y=\epsilon^{j_1\ldots j_{N-1}A}_{i_1\ldots i_{N-1}B}
Y^{i_1}_{j_1} \cdots Y^{i_{N-1}}_{j_{N-1}}
(Z \ldots Z \chi_1 Z\ldots Z\chi_2 Z\ldots Z)^B_A,
\label{ybrane}
\ee
where $\chi_1,\chi_2,\ldots$ represent other SYM fields. 
The other is the $Z=0$ brane open spin chain, 
represented by composite SYM operators 
containing a determinant factor ${\rm det}(Z)$:
\bea
{\cal O}_Z = \epsilon^{j_1 \cdots j_{N-1} A}_{i_1 \cdots i_{N-1} B}
Z_{j_1}^{i_1} \cdots Z_{j_{N-1}}^{i_{N-1}} 
(\chi_{L} Z \cdots Z \chi_1 Z \cdots Z\chi_2 Z \cdots \chi_R)^B_A 
\,\,.\label{zbrane}
\eea
An important difference of the $Z=0$ brane from the $Y=0$ brane is 
that the open SYM 
spin chain is connected to 
the giant graviton 
through boundary 
impurities $\chi_L$ and $\chi_R$. In this note, we are mainly 
interested in the ground states where the bulk magnon 
excitation $\chi_1$, $\chi_2,\cdots$ are absent. 

It is clear that 
the ground state for the $Y=0$ brane is described by 
a unique state because there are no boundary degrees. On the other 
hand, the $Z=0$ brane involves the multiplet of left and right boundary 
states due to the presence of the boundary excitations. Each 
 boundary state is organized by the $SU(2|2)^2$ 
representation. The elementary boundary magnon  involves 16 degenerate 
states with the energy spectrum,
\be
E_B= \sqrt{1+ 4g^2}\,,
\label{energy}
\ee
where $g$ is related to the t' Hooft coupling by
$g= \sqrt{\lambda}/(4\pi)$. 
Each $SU(2|2)$ algebra consists of the $SU(2)\times SU(2)$ rotation 
generators ${\fR}^a\!_b$, ${\fL}^\alpha\!_\beta$, the supersymmetry
generators ${\fQ}^\alpha_a$ and ${\fS}^a_\alpha$ and the central 
charge $\mathfrak{C}$\cite{Beisert:2005tm}. 
 
Their commutators are  given by\cite{Beisert:2005tm}
\bea
&& [{\fR}^a\!_b, \,\, \mathfrak{J}^c]= \delta^c_b \,  
\mathfrak{J}^a-{1\over 2} 
\delta^a_b \, \mathfrak{J}^c\,,\ \ \ 
[{\fL}^\alpha\!_\beta, \,\, \mathfrak{J}^\gamma]= \delta^\gamma_\beta  \, 
\mathfrak{J}^\alpha-{1\over 2} 
\delta^\alpha_\beta \, \mathfrak{J}^\gamma\,
\nonumber\\
&& \{ {\fQ}^\alpha_a, \,\, \fS^b_\beta\}= \delta^b_a   {\fL}^\alpha\!_\beta
+ \delta^\alpha_\beta \, {\fR}^b\!_a
+ \delta^b_a  \delta^\alpha_\beta \mathfrak{C}\,.
\eea 
The central element $\mathfrak{C}$ is related to the energy
by $E_B=2 \mathfrak{C}$. 
There are further central extensions,
\be
\{ {\fQ}^\alpha_a, \,\, \fQ^\beta_b\}=\epsilon^{\alpha\beta}
\epsilon_{ab} {k\over 2}\,,\ \ \ 
\{ {\fS}^a_\alpha, \,\, \fS_\beta^b\}=\epsilon_{\alpha\beta}
\epsilon^{ab} {k^*\over 2}\,.
\ee
For the construction of the boundary states\cite{Hofman:2007xp}, 
we first represent
the $SU(2|2)$ acting on the $2|2$ space. 
We label the bosonic states by
$|\phi^a \rangle$ and the fermionic states by
$|\psi^\alpha \rangle$. Then the generators are acting on the 
states by
\bea
\fR^a\!_b |\phi^c \rangle =
\delta^c_b  |\phi^a \rangle 
-{1\over 2} \delta^a_b  |\phi^c\rangle
\,, \ \ \ 
\fL^\alpha\!_\beta |\phi^\gamma \rangle =
\delta^\gamma_\beta  |\phi^\alpha \rangle 
-{1\over 2} \delta^\alpha_\beta  |\phi^\gamma\rangle
\eea
and by
\bea
&&\fQ^\alpha_a |\phi^b \rangle =
a_B \delta_a^b  |\psi^\alpha \rangle \,, \ \ \ \
\fQ^\alpha_a |\psi^\beta \rangle =
b_B \epsilon^{\alpha\beta}\epsilon_{ab}  |\phi^b \rangle\nonumber\\
&&\fS^a_\alpha|\phi^b \rangle =
c_B \epsilon_{\alpha\beta}\epsilon^{ab} |\psi^\beta \rangle \,, \ \ \ \
\fS^a_\alpha |\psi^\beta \rangle =
d_B \delta^\beta_{\alpha}  |\phi^a \rangle\,.
\eea
The condition
\be 
a_Bd_B-b_B c_B=1
\ee
is necessary for the closure of the algebra. 
One finds also 
that 
$k/2 = a_B b_B$, 
$k^*/2 = c_Bd_B$ and the energy 
$E_B = a_Bd_B+b_B c_B$. From the string theory picture 
explained in \cite{Hofman:2007xp}, 
we assume
that
\be
|k|^2 = 4 g^2\,.
\ee 
Then $a_B$, $b_B$, $c_B$ and $d_B$ are in general parametrized by
\bea
&& a_B =\sqrt{g} \eta_B\,, \ \ \ \ b_B =
{\sqrt{g} f_B \over \eta_B}\nonumber\\
&& c_B =  {i\sqrt{g} \eta_B\over x_B f_B}\,, \ \ \ \ d_B =
{\sqrt{g} x_B \over i\eta_B}\,.
\eea
The unitarity 
demands that 
$f_B$ should be a pure phase with  $|\eta_B|^2 = -i x_B$. The shortening 
condition, $a_Bd_B-b_Bc_B=1$, implies
\be
x_B + {1\over x_B} ={i\over g}\,,\ \ \ \ \ 
x_B = {i\over 2g}\left(1+ \sqrt{1+4g^2}\right)\,. 
\ee
and we recover (\ref{energy}) with
 $E_B ={g\over i}(x_B - x^{-1}_B)$.

Let us denote a representation of one $SU(2|2)$ by $|q_L \rangle =
(|\phi_a \rangle,|\psi_\alpha \rangle)$ with $L=1,2,3,4$ and
$a,\,\, \alpha=1,2$.
 Then the representation of $SU(2|2)\times SU(2|2)$
 is given by the tensor product
 $|q_L\rangle \otimes |q_M\rangle$ corresponding to
the sixteen states. Therefore there are total 256 states
if the both boundary magnons are elementary.


\section{Open string description of boundary giant magnons}
The main purpose of this note is to reproduce
the above ground state degeneracy by studying the open 
string zero mode dynamics.
  
For this purpose,  we shall first describe
the finite-size string ending on the $Z=0$ or the $Y=0$ giant 
gravitons. 
Since we 
are interested in
the boundary degrees,
we focus on the strings without turning on
bulk magnon excitations.

To get the string configuration, 
we first find the classical solution of boundary magnons
with finite size $J$ \footnote{This solution is 
first found in the Ref.~\cite{Note}. The following is the review of the 
solution.}.  
We  begin with the bosonic part of the 
 string action in the conformal gauge,
\be
S = -{\sqrt{\lambda}\over 4 \pi}
\int d\tau \int^{2r}_{0} d \sigma \,\,\,
\partial_a X_I \,\,\partial^a X_I
\label{actionb}
\ee
with the constraint $X_I X_I=1$ ($I=1,2, \cdots, 6$).
We have set the AdS
radial coordinate to zero
since we are interested in the
string moving in $S^5$
while staying at the center of the AdS$_5$. We shall work in
the gauge, $T=\tau$, where $T$
is the global AdS time. 
In this set-up,
the Virasoro constraints
\bea
(\dot{X}_I\pm X'_I)(\dot{X}_I\pm X'_I)= 1\,,
\eea
have to be imposed in addition.

The energy density is uniform in the static  gauge
and the string energy is  proportional to
the spatial coordinate size:
\be
E= {\sqrt{\lambda}\over 2 \pi}\,\, 2r\,.
\ee
For the description of the $Z=0$ or the $Y=0$ boundary states, 
we  turn on 
only $X_1, X_2$ and $X_3$ and
use the coordinates
$Z= X_1 + i X_2 = \sqrt{1-z^2} \,\, e^{i\phi}$ and $X_3 =z$. 

\subsection{String ending on the $Z=0$ brane}

For the $Z=0$ brane, we begin with an ansatz ($\omega \ge 1$),
\be
z= z(\sigma - v \omega \tau)\,,\, \ \ \ \
\phi= \omega \tau + \varphi(\sigma - v \omega \tau)\,.
\ee
The equations of motion are reduced to
\bea
&& (z')^2 = {\omega^2\over (1- v^2\omega^2)^2} \left(z^2 - 1+ {1\over \omega^2}
\right) \left(1-v^2 -z^2\right)\nonumber\\
&& \varphi' = {v \omega^2\over (1- v^2\omega^2)} {z^2 - 1+ {1\over \omega^2}
\over 1-z^2}\,.
\label{eom}
\eea
The general solution can be found as\cite{Zamaklar}
\be
z= {\sqrt{1-v^2} \over \omega \sqrt{\eta}} {\rm dn}\left( {\sigma- v \tau\over
\sqrt{\eta}\sqrt{1-v^2}}\,, \eta
\right)
\ee
where ${\rm dn}(
x, k^2)$ is the Jacobi elliptic function and
 we introduce the parameter $\eta$ by
\be
\eta = {1-\omega^2 v^2\over \omega^2 (1-v^2)}\,.
\ee

Since we are only interested  in the boundary degrees
which are not in motion,
we set $v=0$. The solution then becomes
\bea
&& z= {\rm dn}\left( \omega\, (\sigma-\sigma_0)\,,{1\over \omega^2}
\right)\,,\nonumber\\
&& \phi = \omega\, \tau\,.
\eea
Or in terms of $\sin \theta\equiv \sqrt{1-z^2}$, the solution
is
\be
\sin\theta = {1\over \omega} \,\, {\rm sn}
\left( \omega \, (\sigma-\sigma_0)\,,{1\over \omega^2}\right)\,.
\ee
For the string ending on the $Z=0$ branes,
the Dirichlet boundary condition $\dot{Z}=0$ at $Z=0$
will be imposed for the open string boundaries,
$\sigma=0$ and $\sigma= 2\,r$.
The $\sigma=0$ boundary condition can be   satisfied
 by setting $\sigma_0=0$. The remaining boundary condition
at $\sigma = 2 r $ is satisfied by the choice
$\omega\, r= K(k)$
with $k={1\over \omega}$ where the complete elliptic integrals
$K(k)$ and $E(k)$ are defined by
\bea
&& K(k)=\int_0^1 dx {1\over \sqrt{1-x^2}\sqrt{1-k^2 x^2}}\nonumber\\
&& E(k)=\int_0^1 dx {\sqrt{1-k^2 x^2}\over \sqrt{1-x^2}}\,.
\eea
The other choice $\omega \, r=m K(k) \  (m\  \in\  \mathbb{Z})$ with $m \ge 2$
is possible but it  simply describes the multiple
open strings.

The angular momentum on the $1-2$ plane is given by
\be
J= {\sqrt{\lambda}\over 2\pi} \int^{2r}_0 d \sigma (1-z^2)\dot{\phi}\,.
\ee
Hence the combination $E-J$ can be expanded to the leading correction
for the large $J$ by
\be
E-J = {\sqrt{\lambda}\over \pi} \left(
1- {4\over e^2} e^{- {2 \pi J\over \sqrt{\lambda}}}
+\cdots \right)
=4 g \left(
1- {4\over e^2} e^{- { J\over 2g}}
+\cdots \right)\,.
\label{string}
\ee
The energy and correction is doubled here because we add up
the energy of the left and the right boundary together.
Thus one boundary energy and correction is just one half
of the above $E-J$, which precisely reproduces
the classical part of the energy of (\ref{energy}).
 From the  view point of  classical string,
the construction of the boundary states requires the study
of fermion zero modes. This will lead to
the 16 boundary states for each boundary, so there are
  256 combinations of boundary states in total as we shall see
later on.
Since this construction is independent of the above finite
size correction, we conclude that, for any combination of
boundary states of the left and right boundaries, the
finite energy correction remains the same.

\subsection{Strings ending on the $Y=0$ brane}

In this section, we  describe the
strings ending on the $Y=0$ brane.

We are interested in the open strings moving on the
$Z$ space. Hence at the open string boundaries, one has to
satisfy the Neumann boundary condition
$Z'=0$ since $Z$ is now parallel to the worldvolume of the brane.
The action and the equation of motion are the same as
 (\ref{actionb}) and (\ref{eom}). The trivial solution,
\be
z= \sqrt{1-{1\over \omega^2}}\,,\ \ \ \ \phi= \omega \tau\,,
\ee
satisfies the necessary boundary condition. There is no restriction of
$r$ and $\omega \ge 1$.
The energy
and angular momentum are given by
\bea
E= 4g r\,, \ \ \ \ \ \  J= {4g r \over \omega}\,.
\eea
When $\omega=1$,
one has
\be
E_B=E-J=0
\label{y=0}\,,
\ee
 which corresponds to the
ground state.
It describes a point-like open string carrying finite angular momentum
$J$
moving along the equator\footnote{The  size in the target space is 
point-like here. The finite-size means a finite $J$ corresponding 
to the finite R-charge of SYM spin chain.}.
It is clear that there is no finite size correction
to the energy at least classically.
This is also quite consistent with the fact
 the
$Y=0$ open string does not involve any
 boundary degrees.

\section{Fermion zero modes}
In this section we consider the fermion zero modes
of the strings around the solutions constructed in the previous 
section.
For this we begin with the fermionic part of the 
string action to the quadratic order\cite{Metsaev},
\be
I_F = 2 g \int d\tau d \sigma \,\, L_F\,,
\ee
with
\be
L_F =i(\eta^{ab}\delta_{IJ}- \epsilon^{ab}s_{IJ}) \bar{\theta}^I
\rho_a {D}_b \theta^J\,.
\ee
Here $I$ and $J$ run over $1,2$ and $s_{IJ}$ is diagonal
with $s_{11}=-s_{22}=1$. $\rho_a$ is the world sheet gamma
matrix defined by
\be
\rho_a = \Gamma_A e^A_a = \Gamma_A E^A_\mu \partial_a X^\mu 
\ee
where $\Gamma_A$ and $E^A_\mu$ are respectively 
the 10d gamma matrices, which are taken to be real, and the einbein. $\theta^I$ denotes 
16 component Majonara spinor.
The covariant derivative is defined as
\be
D_a \theta^I  =
(\delta^{IJ} {\cal D}_a -{i\over 2}\epsilon^{IJ}\Gamma_* \rho_a )\theta^J
\ee
where 
\be
{\cal D}_a = \partial_a +{1\over 4} \omega^{AB}_\mu \partial_a X^\mu
 \Gamma_{AB}\,,\ \ \ \ \ \Gamma_* =  i\, \Gamma_{01234}\,.
\ee
The equations of motion take the form,
\bea
&& (\rho_0 -\rho_1)(D_0 + D_1)\theta^1=0\,,\nonumber\\
&& (\rho_0 +\rho_1)(D_0 - D_1)\theta^2=0\,.
\eea 
Let us now work out how the equations look in the background 
we consider. We note that we turn on only $\theta$
and $\phi$ components.
The relevant nonvanishing component of the spin connection
$\omega^{AB}_\mu $ 
is
\be
\omega^{\hat{\phi}\hat{\theta}}_\phi =\cos\theta\,, 
\ee
and
\be
\rho_0 = \Gamma_0 + \omega \sin\theta\Gamma_\phi\,, \ \ \ \
\rho_1 = \phi' \sin\theta\Gamma_\phi +
\theta' \Gamma_\theta =\theta' \Gamma_\theta  \,,
\ee
where we have used $\dot\theta=0$,  $\dot\phi=\omega\,\,\,$ 
and $\phi'=0$.
Therefore the equations become
\bea
&& (\rho_0 -\rho_1)\Big[\partial_t \theta^1 + D\theta^1 -
{i\over 2} \Gamma_* (\rho_0 + \rho_1)\theta^2\Big]=0\nonumber\\
&& (\rho_0 +\rho_1)\Big[-\partial_t \theta^2 + \bar{D}\theta^2 -
{i\over 2} \Gamma_* (\rho_0 - \rho_1)\theta^1\Big]=0\,,
\eea
where we introduce
\be
D=\partial_\sigma +{\omega\cos\theta\over 2}\Gamma_{\phi\theta}\,,\ \ \ \ 
\bar{D}=\partial_\sigma -{\omega\cos\theta\over 2}\Gamma_{\phi\theta}\,.
\ee
These equations are further rewritten as
\bea
&& (\partial_t  + D)\psi^1 -
{i\over 2}[\rho_0,\, \Gamma_*]\psi^2=
(\partial_t  + D)\psi^1 +
{i}\omega \sin\theta \Gamma_*\Gamma_\phi\psi^2=0
\nonumber\\
&& 
(-\partial_t  + \bar{D})\psi^2 -
{i\over 2} 
[\rho_0, \,\Gamma_*]
\psi^1=
(-\partial_t  + \bar{D})\psi^2 +
{i} 
\omega \sin\theta \Gamma_*\Gamma_\phi
\psi^1
=0
\,,
\eea
where we introduced new spinors $\psi^I$ defined by
\be
\psi^1=
i (\rho_0 -\rho_1)\theta^1\,,\ \ \ \ 
\psi^2=
i (\rho_0 +\rho_1)\theta^2\,,
\ee
and used the relations
\bea
&&[\rho_0,\, \Gamma_*]= -2\omega\, \sin\theta\,\Gamma_* \Gamma_\phi\,,\\
&& [\rho_0-\rho_1, \, D]=[\rho_0+\rho_1\,, \bar{D}]=0\,.
\eea
The boundary contributions of the variation of the action should vanish,
which leads to the condition,
\be
\Big[\bar{\theta}^1(\rho_0-\rho_1)\delta\theta^1 -
\bar{\theta}^2(\rho_0+\rho_1)\delta\theta^2\Big]_{boundary}=0\,.
\label{boundary}
\ee 

Finally for the zero mode, the equations are reduced to
\bea
&& D\, \psi^1 +
{i}\omega \sin\theta \Gamma_*\Gamma_\phi\psi^2=0
\nonumber\\
&& 
\bar{D}\psi^2 +
{i} 
\omega \sin\theta \Gamma_*\Gamma_\phi \,
\psi^1
=0
\,.
\eea
By eliminating $\psi^2$, one gets
\be
\Big[\Big({1\over \omega \sin\theta}\,  D \Big)^2 -1\Big]\psi^1=0\,,
\ee
which is equivalent to two first-order equations,
\be
D\psi^1 =\mp \omega \sin\theta\, \psi^1\,.
\label{master}
\ee
Then $\psi^2$ is given by
\be
\psi^2 = \pm i\, \Gamma_* \Gamma_\phi \,\psi^1\,. 
\label{relation}
\ee
One comment is that the boundary condition 
 (\ref{boundary})
is satisfied  automatically
since the boundary term vanishes   
with the relations (\ref{relation}).
These equations will be the starting point of our analysis 
of the fermion zero modes.

\subsection{Zero modes for the $Z=0$ boundary}

For the $Z=0$ boundary, we use
$\cos\theta= {\rm dn}(\omega\sigma)$ and $\sin\theta= k\,{\rm sn}(\omega
\sigma)$.
The solution of (\ref{master}) can be found as
\bea
\psi_\pm^1 &=& 
i \, N(k)\, (\rho_0-\rho_1)\big[{\rm dn}(\omega \sigma) 
\pm k \, {\rm cn}(\omega \sigma)
\big]
\big[{\rm sn}(\omega\sigma)
+ {\rm cn}(\omega\sigma)
\Gamma_{\phi\theta}\big]^{1\over 2}\, U_\pm\nonumber\\
&=& i\, N(k) \,\big[{\rm dn}(\omega \sigma) 
\pm k \, {\rm cn}(\omega \sigma)
\big]
\big[{\rm sn}(\omega\sigma)
+ {\rm cn}(\omega\sigma)
\Gamma_{\phi\theta}\big]^{1\over 2} \big[\Gamma_0 
+\Gamma_\phi 
\big]\, U_\pm\,,
\eea 
where $U_\pm$ is a constant Majonara spinor and 
the normalization factor $N(k)$ defined by 
\be
N^2(k)={1 \over 2k(1+k)E\Big({2\sqrt{k}\over 1+k }\Big)}
\ee
is introduced for the 
normalization.
Note that $D\psi^1|_{\rm boundary} = \bar{D}\psi^2|_{\rm boundary}=0$
and we shall require these as extra boundary 
conditions for the $Z=0$ and the $Y=0$ branes.

The solutions $\psi^I_+$ and $\psi^I_-$ have a maximum
at $\sigma=0$ and $\sigma=2 k\, K(k)$ respectively. 
Hence $\psi^I_+$ is concentrated 
on the left boundary $\sigma=0$
while $\psi^I_-$ is concentrated on the right 
boundary $\sigma=2 k\, K(k)$. Therefore $\psi^I_+$ and 
$\psi^I_-$ can be viewed
as describing the left and the 
right boundary degrees respectively. 
This becomes clear if the string length $2k\, K(k)$ 
becomes infinite as $k\rightarrow 1$.    
For $k\rightarrow 1$, the Jacobi elliptic functions become
\bea
&&{\rm sn}(\omega\sigma)\ \rightarrow\  \tanh \sigma \nonumber\\
&& {\rm dn} 
 (\omega\sigma)\,,\   {\rm cn} (\omega\sigma)
\ \rightarrow\ 
{1/ \cosh\sigma}\,.
\eea
Also for $k\rightarrow 1$, one has 
\bea
&&{\rm sn}(\omega\sigma)\ \rightarrow\  \tanh (2r-\sigma) \nonumber\\
&&   {\rm dn} (\omega\sigma)\,, \ -{\rm cn} (\omega\sigma) \ \rightarrow\ 
{1/ \cosh(2r-\sigma)}\,,
\eea
when $2r-\sigma$ is finite. In this limit, one can check that the 
overlap of $\psi^I_+$ and $\psi^I_-$ disappear completely. 

The full 
solution is
 given by
\be
\psi^1= \psi^1_+ +\psi^1_-\,, \ \ \ \ \ \ 
\psi^2= i\, \Gamma_* \Gamma_\phi (\psi^1_+ -\psi^1_-)\,.
\ee
The effective action for the zero mode dynamics 
can be obtained by giving the time dependence
to the zero mode fermion coordinates as $U_\pm(\tau)$; This leads to
\bea
I_{\rm zero} =
2g \
\int d\tau \Big[ i U^T_+ (\Gamma_0+\Gamma_\phi)^\dagger (\Gamma_0+\Gamma_\phi) 
\dot{U}_+ 
+i U^T_- (\Gamma_0+\Gamma_\phi)^\dagger (\Gamma_0+\Gamma_\phi) 
\dot{U}_-\Big]\,,
\eea
where the cross terms from $\psi^1$ and $\psi^2$ are cancelling with each 
other. 
With further definitions,
\bea
 U_L = (\Gamma_0+\Gamma_\phi ) \,\, U_+ \,,  \ \ \ 
U_R =  (\Gamma_0+\Gamma_\phi )\,\, U_- \,,
\eea
the action for the zero mode dynamics becomes
\be       
I_{\rm zero} =
2g \,\,
\int d\tau \Big[ i\, U_L^T 
\dot{U}_L 
+ i\, U^T_R
\dot{U}_R
\Big]\,.
\label{action}
\ee 
It is clear that the left and the right degrees behave 
independently. Let us consider the dynamics of the left boundary 
first. Among the 16 real components $U_L$, the light-con condition
$(\Gamma_0+\Gamma_\phi)\,\, U_L=0$ projects down by half and only 8 real 
degrees remain. We organize this  in terms of the bispinor
components
$U_{\alpha a}$ and $\tilde{U}_{\dot{\alpha}\dot{a}}$ where
$\alpha$ and $\dot{\alpha}$ are the spinor indices
for $SO(4)\,\,\simeq \,\, SU(2)\times SU(2)$ isometry in the transverse part of
AdS$_5$
and
the $a$ and $\dot{a}$ 
for $SO(4)$ isometry  in the transverse part of
$S^5$\cite{Minahan:2007gf}.
The quantization leads to the anticommutation relations
\bea
&& \{ U_{\alpha a}\,, U_{\beta b}\}=
{1\over 2g}\epsilon_{\alpha\beta}\epsilon_{ab}
\nonumber\\
&& \{ \tilde{U}_{\dot\alpha \dot{a}}\,, \tilde{U}_{\dot\beta \dot{b}}\}=
{1\over 2g}\epsilon_{\dot\alpha\dot\beta}\epsilon_{\dot{a}\dot{b}}
\nonumber\\
&&  \{ U_{\alpha a}\,, \tilde{U}_{\dot\beta \dot{b}}\}=
0 \,.
\eea
Then the Hilbert space and operators are realized as
\bea
&& U_{\alpha a }|\phi_b\rangle ={1\over \sqrt{2g}} \epsilon_{ba}|\psi_\alpha
\rangle\nonumber\\
&&  U_{\alpha a }|\psi_\beta \rangle ={1\over \sqrt{2g}} 
\epsilon_{\alpha\beta}|\phi_a
\rangle
\eea
up to the freedom of the usual unitary transformations.
Then the operators
$\fQ_{\alpha a}$ and  $\fS_{\alpha a}$ are realized as
\bea
&& \fQ_{\alpha a } ={ \sqrt{g\over 2}}\Big[
(a_B + b_B)-(a_B - b_B)(-)^F
\Big] 
U_{\alpha a }
\nonumber\\
&&  \fS_{\alpha a } ={\sqrt{g\over 2}} 
\Big[
(c_B + d_B)-(c_B - d_B)(-)^F
\Big] 
U_{\alpha a }
\eea
where $F$ is the fermion number operator with
 $(-)^F 
U_{\alpha a } + 
U_{\alpha a }(-)^F=0$.

For the $\tilde{U}$, one has the same construction and there are consequently
altogether 
16 states for the left boundary. For the right boundary, one has 16 
states too.
Therefore the 256 states with the precise algebra are constructed, which is 
matching with the Yang-Mills theory side. 

\subsection{Zero modes for the $Y=0$ boundary}
For the  
$Y=0$ boundary, we use the vacuum solution, $\cos\theta=1$ and $\phi=\tau$.
The equations in (\ref{master}) become
\be
{d\over d\sigma}\,\,\psi^1= \pm\,\, \psi^1\,.
\ee 
Then the most general solution is
\be
\psi^1 =i\,(\Gamma_0+\Gamma_\phi) \Big[ e^{-\sigma}\, U_+
+ e^{-(2r-\sigma)}\, U_-\,
\Big]\,.
\ee
However there is no way to satisfy the boundary 
conditions\footnote{The Dirichlet type boundary condition 
$\psi^I(0)= \psi^I(2r)=0$ 
does not allow any zero modes either. 
One needs a further clarification of the required 
boundary conditions. This is an interesting problem but beyond the 
scope 
of this paper. The author would like to thank A. Tseytlin
for his comment on this point.} 
$D\psi^1|_{\rm boundary} = \bar{D}\psi^2|_{\rm boundary}=0$. 
Hence there are no fermion zero 
modes. Therefore the corresponding state is simply 
describing the 
unique vacuum,
which is again consistent with the 
SYM  theory construction of the state.

\section{Discussions}

In this paper, we studied the fermion zero mode problem
for the open strings ending on the $Z=0$ or the $Y=0$ branes. 
For the open string ending on the $Z=0$ brane, the fermion zero mode dynamics 
reproduces the 256 states of the boundary degrees with the 
precise realization 
of the $SU(2|2)\times SU(2|2)$ symmetry algebra.
Also for the open string ending on the $Y=0$ brane, we reproduce
the unique vacuum state by studying its fermion zero-mode dynamics.

Recently there appeared the proposal of the correspondence
between the ${\cal N}=6$ 
super Chern-Simons theory and 
the strings on the AdS$_4\times \mathbb{CP}\,3$\cite{Aharony:2008ug}. 
The integrability of the  planar two loop 
integrability is checked in Refs.\cite{Mina,Bak} with  further 
developments\cite{Gaio}.
The string sigma model, its giant magnon solutions and some related issue
 are also studied in 
Refs.~\cite{Arut,Grig, Kri}.  
However we still lack some direct information on  
the states of the elementary magnon of this Chern-Simons 
spin chains.
In this sense, the identification of the 
states with symmetry algebra arising 
from the fermion zero-mode dynamics is of much interest.

\section*{Acknowledgement}
We are grateful to C. Ahn,  S.-J. Rey and A. Tseytlin 
for many helpful discussions. 
This work was supported in part
by UOS 2007 Academic Research Grant.


\end{document}